\documentclass[12pt,a4paper]{article}

\title {On the hydrino state of the relativistic hydrogen atom}
\author {Jan Naudts\\
  Departement Natuurkunde, Universiteit Antwerpen,\\
  Universiteitsplein 1, 2610 Antwerpen, Belgium\\
  E-mail {\tt Jan.Naudts@ua.ac.be}
}

\usepackage{amsfonts}
\usepackage{amsmath}

\def\Ro{{\mathbb R}}

\def\qq{{\bf q}}

\def\angstrom{\hbox{\AA}}

\newcommand{\be}{\begin{eqnarray}}
\newcommand{\ee}{\end{eqnarray}}

\begin {document}
\date {}

\maketitle 

\begin {abstract}
The Klein-Gordon equation of the hydrogen atom has a low-lying eigenstate,
called hydrino state, with square integrable wavefunction. The corresponding spinor solution
of Dirac's equation is not square integrable. For this reason the hydrino state has been rejected
in the early days of quantum mechanics as being unphysical.
Maybe it is time to change opinion.
\end {abstract}

%%%%%%%%%%%%%%%%%%%%%%%%%%%%%%%%%%%%%%%%%%%%%%%%%%%%%%%%%%%%%%%%%%%%%%%%%%%%%%
%%%%%%%%%%%%%%%%%%%%%%%%%%%%%%%%%%%%%%%%%%%%%%%%%%%%%%%%%%%%%%%%%%%%%%%%%%%%%%
\section {Introduction}

A.~Sommerfeld \cite {SA23} was first to calculate relativistic corrections to
the energy levels of the hydrogen atom, even before Heisenberg and Schr\"odinger
laid the foundations of non-relativistic quantum mechanics. Later the energy levels
were calculated using the relativistic equations of Dirac. The result happened to
coincide in leading order with that of Sommerfeld. These calculations are repeated in about every
textbook on relativistic quantum mechanics --- see e.g.~Sakurai \cite {SJJ67}.

R.L.~Mills and collaborators recently \cite {MR02,MNR02,MR03,MRDMH,PMC04} discovered
the formation of plasma in a hydrogen gas under rather exceptional conditions.
They explain the observed phenomena by a catalytic conversion of hydrogen atoms
into a state where the electron is strongly bound to the nucleus, with a binding
energy which is much larger than in the conventional groundstate of the hydrogen atom.
They call this new state the hydrino state.
A. Rathke \cite {RA05} has questioned the existence of such a state,
claiming that it is incompatible with standard quantum mechanics.
All Rathke's arguments relate to non-relativistic quantum mechanics.
The present paper discusses the problem in the context of relativistic
quantum mechanics.

In the next Section the groundstate of the relativistic Klein-Gordon equation is considered.
Section 3 deals with spherically symmetric excited states. In Section 4 
an argument is given, explaining how it is possible that the hydrino solution of the
Klein-Gordon equation is square integrable, while the corresponding solution
of the Dirac equation is not.
The final section contains a short discussion.

%%%%%%%%%%%%%%%%%%%%%%%%%%%%%%%%%%%%%%%%%%%%%%%%%%%%%%%%%%%%%%%%%%%%%%%%%%%%%%
%%%%%%%%%%%%%%%%%%%%%%%%%%%%%%%%%%%%%%%%%%%%%%%%%%%%%%%%%%%%%%%%%%%%%%%%%%%%%%
\section {Relativistic equation}

Replace the time-dependent Schr\"odinger equation
\be
i\hbar\frac {\partial\,}{\partial t}\psi_t=-\frac {\hbar^2}{2m_0}\Delta\psi_t+V\psi_t
\ee
by the relativistic equation of motion
\be
\left(i\hbar\frac {\partial\,}{\partial t}-V\right)^2\psi_t
+\hbar^2c^2\Delta\psi_t
=c^4m_0^2\psi_t.
\label {releq}
\ee
This is the Klein-Gordon equation, with minimal coupling to a non-quantised static (electromagnetic) field.
Solutions of this equation are relativistic wavefunctions and should be square integrable functions of Euclidean space $\Ro^3$.

Consider now the Coulomb potential
\be
V(q)=-\alpha\frac {\hbar c}r
\qquad\hbox{ with }r=|q|
\ee
and try a stationary  solution of the form
\be
\psi_t(q)=e^{-i\hbar^{-1}E_0t}r^{-l}e^{-r/r_0}
\ee
with $l<3/2$ and $r_0>0$. These conditions are needed to guarantee square integrability of the wavefunction.

Using
\be
\Delta\psi_t&=&\frac 1{r^2}\frac{\partial\,}{\partial r}r^2\frac{\partial\,}{\partial r}\psi_t\cr
&=&\frac 1{r_0^2}\left(-l(1-l)\frac {r_0^2}{r^2}-2(1-l)\frac {r_0}{r}+1\right)\psi_t
\ee
the equation becomes
\be
\left(E_0+\alpha\frac {\hbar c}r\right)^2
+\frac {\hbar^2c^2}{r_0^2}\left(-l(1-l)\frac {r_0^2}{r^2}-2(1-l)\frac {r_0}{r}+ 1\right)
=c^4m_0^2.
\ee
Collecting terms in the same power of $r$ yields the set of equations
\be
E_0^2+\frac {\hbar^2c^2}{r_0^2}&=&c^4m_0^2
\label {gseq1}\\
E_0\alpha&=&(1-l)\frac {\hbar c}{r_0}
\label {gseq2}\\
\alpha^2&=&l(1-l).
\label {gseq3}
\ee
The last equation has two solutions (note that $\alpha\simeq 1/137<1/2$)
\be
l=\frac 12\left(1\pm\sqrt{1-4\alpha^2}\right).
\ee
The two remaining equations can be written as
\be
E_0&=&\gamma m_0c^2
\label {Enotsol}\\
r_0&=&\frac {\hbar}{m_0c}\frac 1{\sqrt l}
\label {rnotsol}
\ee
with
\be
\gamma=\frac{\frac {\alpha}{l}}{\sqrt{1+\frac {\alpha^2}{l^2}}}=\frac {\alpha}{\sqrt l}=\sqrt{1-l}.
\ee
One of the two solutions gives a small value of $l$. Then $\alpha/l\simeq 1/\alpha$ so that
the above equations become
\be
E_0&\simeq&m_0c^2(1-\frac 12\alpha^2+\cdots)\\
r_0&\simeq&\frac {\hbar}{m_0c}\cdot\frac 1\alpha\simeq 0.53\angstrom.
\ee
The other solution gives an $l$-value close to 1. Then the results become
\be
E_0&\simeq&m_0c^2\cdot\alpha\\
r_0&\simeq&\frac {\hbar}{m_0c}\simeq 0.0039\angstrom.
\ee
In the latter case the electron is strongly bounded to the proton, almost $10^4$ times stronger than in the
conventional ground state of the electron.
Such a state, if it exists, may be called a hydrino state \cite {MRL99,RA05}.
Note that the radius of the hydrino state, as found here, is to leading order independent of the actual value of $\alpha$.

In textbooks (see e.g.~\cite {SJJ67}), the solution with $l$ close to 1 is rejected as being non-physical.
This point will be discussed later on.

%%%%%%%%%%%%%%%%%%%%%%%%%%%%%%%%%%%%%%%%%%%%%%%%%%%%%%%%%%%%%%%%%%%%%%%%%%%%%%
%%%%%%%%%%%%%%%%%%%%%%%%%%%%%%%%%%%%%%%%%%%%%%%%%%%%%%%%%%%%%%%%%%%%%%%%%%%%%%
\section {Excited states}

In this section excited state solutions of the Klein-Gordon equation (\ref {releq})
are calculated. For simplicity, only spherically symmetric states are considered.
Make the {\sl ansatz}
\be
\psi_t(\qq)=e^{-i\hbar^{-1}E_nt}r^{-l}e^{-r/r_n}p_n\left(r/{r_n}\right)
\ee
with $p_n(x)$ a polynomial of degree $n$
\be
p_n(x)=\sum_{j=0}^na_j x^j.
\ee
One calculates
\be
& &\Delta\psi_t(\qq)\cr
&=&\frac 1{r_n^2}
\left[-l(1-l)\frac {r_n^2}{r^2}-2(1-l)\frac{r_n}{r}+1
+\frac {p''}p+2\left((1-l)\frac {r_n}r-1\right)\frac {p'}p
\right]\psi_t(\qq).\cr
& &
\ee
Hence the equation becomes
\be
& &\left(E_n+\frac {\alpha\hbar c}r\right)^2\cr
&+&
\frac {\hbar^2 c^2}{r_n^2}
\left[-l(1-l)\frac {r_n^2}{r^2}-2(1-l)\frac{r_n}{r}+1
+\frac {p''}p+2\left((1-l)\frac {r_n}r-1\right)\frac {p'}p
\right]\cr
&=&c^4m_0^2.\cr
& &
\ee
After multiplication with $p_n$ this becomes
\be
& &
p_n''+2\left((1-l)\frac {r_n}r-1\right)p_n'\cr
&+&\left[\frac {r_n^2}{\hbar^2 c^2}\left(E_n+\frac {\alpha\hbar c}r\right)^2
-l(1-l)\frac {r_n^2}{r^2}-2(1-l)\frac{r_n}{r}+1
-\frac {c^2m_0^2r_n^2}{\hbar^2}\right]p_n\cr
&=&0.\cr
& &
\ee
Equating the terms in $r^n$, resp.~$r^{-2}$,  gives condition (\ref {gseq1}), resp.~(\ref {gseq3}),
with $E_0$ replaced by $E_n$ and $r_0$ replaced by $r_n$.
From the terms in $r^{n-1}$ follows
\be
E_n\alpha=\frac {\hbar c}{r_n}(n+1-l).
\ee
This equation replaces (\ref {gseq2}).

\begin{table}

\begin{tabular}{|l|l|l|}
\hline
$E_0/m_0c^2=\gamma$  &0.999973372546	  &0.007297547395  \\ 
\hline
$E_1/m_0c^2$  &0.999993343292	  &0.999973378218  \\ 
\hline
$E_2/m_0c^2$  &0.999997041499	  &0.999993344001  \\ 
\hline 
\end{tabular} 

\caption  {Lowest eigenvalues,
according to (\ref {sommerfeld}), for the two possible values of $\gamma$, given $\alpha=0.00729735308\simeq 1/137$.}
\end{table} 

The solution for the energy value can be written as
\be
E_n&=&m_0c^2\frac {n+\gamma^2}{\sqrt {n^2+(2n+1)\gamma^2}}\cr
&=&\frac {m_0c^2}{\sqrt{1+\frac {\alpha^2}{(n+1-l)^2}}}.
\label {sommerfeld}
\ee
The latter expression was first obtained by A. Sommerfeld \cite {SA23}
and gives the energy levels of the hydrogen atom, with relativistic corrections to leading order in $\alpha$
(see e.g.~\cite {SJJ67}, Eq.~3.311).
It is clear from this expression that the binding energy
is always small compared to $m_0c^2$, except when $n=0$ and
$\gamma$ is small (see Table 1). This one exception is
the only state for which the binding energy is larger than that
of the the traditional ground state. It is also the only one 
for which the Bohr radius, given by
\be
r_n=\frac {\hbar}{m_0c}\frac 1\alpha\sqrt{n^2+(2n+1)\gamma^2},
\ee
is smaller. Hence, the present treatment predicts the existence of a single hydrino state
which is more stable than the conventional groundstate of the hydrogen atom.

%%%%%%%%%%%%%%%%%%%%%%%%%%%%%%%%%%%%%%%%%%%%%%%%%%%%%%%%%%%%%%%%%%%%%%%%%%%%%%
%%%%%%%%%%%%%%%%%%%%%%%%%%%%%%%%%%%%%%%%%%%%%%%%%%%%%%%%%%%%%%%%%%%%%%%%%%%%%%
\section {Radial Dirac equation}
\label {dirac}

Here, we try to understand how it is possible that the Klein-Gordon equation
has a hydrino solution, which is square integrable, while the corresponding
solution of the Dirac equation is not square integrable.
A transition is made via the equation of van der Waerden (see \cite {SJJ67}) to transform a
solution of the Klein-Gordon equation into a spinor solution of Dirac's equation.
The first step of this transformation is to replace the second order equation by a
couple of first order equations.

Let be given a stationary solution $\psi$ of (\ref {releq})
with energy $E$.
Try to find functions $\phi$, 
$u$, $v$, $x$, and $y$ such that
\be
r\frac {{\rm d}\psi}{{\rm d}r}&=&u\psi+v\phi
\label {rd1}\\
r\frac {{\rm d}\phi}{{\rm d}r}&=&x\psi+y\phi.
\label {rd2}
\ee
Eliminating $\phi$ from these equations gives
\be
\psi''+\left(\frac {1-u-y}r-\frac {v'}v\right)\psi'
+\left(-\frac {u'}r+\frac {uy}{r^2}-\frac {vx}{r^2}+\frac {uv'}{rv}\right)\psi=0.
\ee
This equation should coincide with the stationary Klein-Gordon equation
\be
\psi''+\frac 2r\psi'+\left[
\left(\frac E{\hbar c}+\frac {\alpha}r\right)^2-\frac {c^2m_0^2}{\hbar^2}\right]\psi=0.
\ee
Comparison yields a possible solution with $v=-\alpha^{-2}r$ and
\be
u&=&-1-\frac {mc}{\hbar}r\\
y&=&-1+\frac {mc}{\hbar}r\\
x&=&\alpha^2r
\left(\frac E{\hbar c}+\frac {\alpha}r\right)^2.
\ee
With this choice of functions the pair of first order equations (\ref {rd1}, \ref {rd2})
is equivalent with the Klein-Gordon equation (\ref {releq}). 

Assume now that $\psi$ is a solution of the type described in previous sections,
divergent as $r^{-l}$ for small $r$. Then $\phi$ can be calculated via (\ref {rd1})
and diverges as $r^{-l-1}$. Hence it is a square integrable function provided that
$l<1/2$. This condition is satisfied for all of the usual eigenstates of the hydrogen atom,
but not for the hydrino state for which $l$ is close to 1.
This argument shows that the hydrino ground state has a square integrable wavefunction,
which is solution of the Klein-Gordon equation, but the corresponding spinor,
which solves the Dirac equation, is not square integrable. Note that
the equations (\ref {rd1}, \ref {rd2}) are not identical with the radial Dirac equations,
as found in literature,
because external fields are treated slightly different in the Dirac equation
compared with the Klein-Gordon equation. But this difference is not important
for the argument of the present section.

%%%%%%%%%%%%%%%%%%%%%%%%%%%%%%%%%%%%%%%%%%%%%%%%%%%%%%%%%%%%%%%%%%%%%%%%%%%%%%
%%%%%%%%%%%%%%%%%%%%%%%%%%%%%%%%%%%%%%%%%%%%%%%%%%%%%%%%%%%%%%%%%%%%%%%%%%%%%%
\section {Discussion}

This paper starts with the Klein-Gordon equation, with minimal coupling to the
non-quantised electromagnetic field. In case of a Coulomb potential this
equation is the obvious relativistic generalisation of the Schr\"odinger
equation of the non-relativistic hydrogen atom, if spin of the electron is neglected.
It has two sets of eigenfunctions,
one of which introduces small relativistic corrections to the non-relativistic
solutions. The other set of solutions contains one eigenstate which describes
a highly relativistic particle with a binding energy which is a large fraction of
the rest mass energy. This is the hydrino state.

Section \ref {dirac} tries to transform the wavefunction of the hydrino into a spinor solution
of the Dirac equation. During this transformation, components of the spinor are
obtained by taking derivatives of the hydrino wavefunction. As a result, these
components are not any longer square integrable. The existence of an extra set
of solutions of the Dirac equation of the hydrogen atom is known since long.
The fact that these extra solutions are not square integrable
has been used as an argument to reject them as being non-physical.
However, because the corresponding solutions of the Klein-Gordon equation
are square integrable, a number of difficult questions come up.
Is it an axiom of physics that the spinor solutions
of Dirac's equation are square integrable? Or is Dirac's equation just a convenient way
to handle the Klein-Gordon equation in case of a spin-1/2 particle, and is the basic requirement that the
solution of the latter equation is square integrable?

Even if Dirac's equation constitutes the basic theory
then it is still not clear whether non-integrability at the position of the nucleus is more than a
mathematical inconvenience. Indeed, the nucleus of the hydrogen atom is not a point
but its charge is smeared over a distance of about $10^{-15}$ m. The solutions of
the Klein-Gordon equation or of the Dirac equation with smeared-out Coulomb potential
are expected not to diverge at the origin. Hence the problem of square integrability
is not a physical problem. The equations cannot be solved analytically in this case. But there is no reason
why the hydrino solution should disappear. The main effect of smearing out the Coulomb potential,
besides regularisation of the wavefunction at the origin, will be a small decrease of the binding energy.

In a recent paper, Rathke \cite {RA05} criticised the unconventional theory of Mills \cite {MRL99}
concerning the existence of the hydrino state of the hydrogen atom. The
present paper shows that one can find arguments in favour of the hydrino state
also in the standard theory of relativistic quantum mechanics. The model of the hydrogen
atom considered here is not the most sophisticated one. The motion of the nucleus and the spins of electron and of
nucleus have been neglected. The electromagnetic field should be treated in second quantisation.
These modifications will add a lot of technicality but will probably add only minor corrections
to the present treatment. As long as these more sophisticated calculations are not
accomplished, there are no serious arguments from quantum mechanical theory to reject the
existence of the hydrino state.

\section*{}
%%%%%%%%%%%%%%%%%%%%%%%%%%%%%%%%%%%%%%%%%%%%%%%%%%%%%%%%%%%%%%%%%%%%%%%%%%%%%%
%%%%%%%%%%%%%%%%%%%%%%%%%%%%%%%%%%%%%%%%%%%%%%%%%%%%%%%%%%%%%%%%%%%%%%%%%%%%%%

\end {document}